\documentstyle[12pt]{article}

\begin{document}

\newcommand{\beq}{\begin{equation}}
\newcommand{\eeq}{\end{equation}}

\begin{titlepage}

\begin{center}

\hfill hep-th/9701153

\vskip 1.8cm
 {\bf Thermal Action and Specific Heat of \\
 the Five-Dimensional Non-Extremal Black Hole }

\vskip .8cm

Shijong Ryang

\vskip .8cm
{\em Department of Physics \\ Kyoto Prefectural University of Medicine \\
Taishogun, Kyoto 603 Japan}

\end{center}

\vskip 2.8cm

\begin{center} {\bf Abstract} \end{center}
We construct the Euclidean on-shell action for the five-dimensional
non-extremal black hole with multiple electric charges. We show that this
thermal action agrees with one half of the entropy. This agreement is
argued to be related to the generalized Smarr formula of the
five-dimensional black hole mass. Through the calculation of the specific
heat far off extremality we observe that a phase transition occurs.

\vskip 7cm
\noindent January 1997

\end{titlepage}

Recently there has been considerable attention to the study of the black
hole thermodynamics in string theory. The microscopic degeneracies of black
holes were explained from the D-brane bound states. The Bekenstein-Hawking
entropy of a certain extremal five-dimensional Reissner-Nordsr\"{o}m black
hole was computed by counting the asymptotic degeneracy of BPS oscillations
of an effective string that describes the bound state of D-$5$-branes
and D-$1$-branes \cite{SV}. The computations of the entropies were
 further performed for the near-extremal six-dimensional black string
\cite{HS} and the near-extremal five-dimensional black holes \cite{CM,HMS}
with multiple charges, which were shown to be associated with the degeneracy
of low-lying non-BPS oscillations of the effective string. The remarkable
statistical derivations of the entropies were extended to the
four-dimensional  black holes \cite{MS1,JKM} and the rotating
black holes \cite{BMPV,BLMP}.

On the other hand since Gibbons and Hawking \cite{GH} showed that through
the saddle-point approximation the Euclidean partition function for quantum
gravity is interpreted as the thermal partition function for a system of
the black hole temperature, the thermodynamic properties of various black
holes were investigated. Based on such direct calculation of the partition
function the thermodynamics was constructed for the four-dimensional
rotating black holes \cite{BMY} or the four-dimensional U$(1)^2$ dilaton
black holes \cite{KOP}, where the rule that the entropy is one quarter of
the area of the event horizon was discussed. Recently to study the zero
mass black holes the four-dimensional anti-Maxwell dilaton Einstein theory
as well as the usual Maxwell one, that are specified by a dimensionless
dilaton coupling parameter $a$, were considered \cite{GR} and the on-shell
actions for the electrically singly-charged non-extremal black holes
were evaluated to be equal to the entropies. The specific heats
characterized by $a$ were shown to behave differently between both theories.

We will investigate the five-dimensional non-extremal black hole with
multiple electric charges by means of the Gibbons-Hawking method. We will
derive the Euclidean on-shell action and discuss the thermal properties
of this five-dimensional black hole, which will be compared with those
of the four-dimensional charged dilaton black holes. The specific heat
will be computed for the non-extremal black hole as well as the
near-extremal one, from which the occurrence of a phase transition will
be suggested.

We start with the six-dimensional low-energy Lorentzian action for the type
IIB string theory in the canonical frame
\beq
\frac{1}{16{\pi}G_6}\int d^6x \sqrt{-g}\left( R - (\nabla\phi)^2 -
\frac{1}{12}e^{2\phi}H^2 \right),
\eeq
where $H$ is the RR three-form field strength \cite{HS}. Through the black
string solution specified by
\beq
ds_6^2 = e^{2D}(dx_5 + A_{\mu}dx^{\mu})^2 + ds_5^2,
\label{ds}\eeq
where $x^{\mu}$ are the coordinates of the five-dimensional spacetime and
$x_{5}$ is on an $S^1$ of circumference $L$, the action is truncated into
the five-dimensional one
\beq
\frac{1}{16{\pi}G_5}\int d^5x \sqrt{-g}e^D \left( R - (\nabla\phi)^2
- \frac{2}{3}(\nabla D)^2 - \frac{e^{-2D+2\phi}}{4}H_{+}^2 \nonumber \\
- \frac{e^{-2D-2\phi}}{4}H_{-}^2 - \frac{e^{2D}}{4}G^2 \right)
\label{Ac}\eeq
with $G_{5} = G_{6}/L$. We set the six-dimensional Newton constant $G_6 =
1$. This action has three U$(1)$ gauge fields such as the usual
Kaluza-Klein field strength $G = dA, (H_{+})_{\mu\nu} = H_{\mu\nu5}$ and
$H_{-} = e^{2\phi+D}*H$ where $*$ is the five-dimensional Hodge dual.
Here we write down the non-extremal six-dimensional black string solution
in the canonical frame
\begin{eqnarray}
ds_6^2 = -\left[ 1 - \left(\frac{r_{+}^2\cosh^2\alpha -
r_{-}^2\sinh^2\alpha}{r^2}\right) \right]dt^2 + \sinh2\alpha
\frac{r_{+}^2 -r_{-}^2}{r^2}dtdx_5 \nonumber \\ + \left[ 1 - \left(
\frac{r_{-}^2\cosh^2\alpha - r_{+}^2\sinh^2\alpha }{r^2} \right)\right]
dx_5^2 + \left(1 - \frac{r_{-}^2}{r^2}\right)^{-1}\left(1 -
\frac{r_{+}^2}{r^2}\right)^{-1}dr^2 + r^2d\Omega_3^2
\label{dsa}\end{eqnarray}
with a boost parameter $\alpha$, which is constructed by boosting the
non-extremal zero-momentum black string solution found in \cite{GHT},
where the event horizon and the inner horizon are located at $r = r_{+},
 r_{-}$ respectively. Comparison of this expression with (\ref{ds}) leads
to
\beq
e^{2D} = 1 - \frac{r_{-}^2\cosh^2\alpha - r_{+}^2\sinh^2\alpha}
{r^2} ,\; A_t =
e^{-2D}\sinh2\alpha\frac{r_{+}^2 - r_{-}^2}{2r^2}.
\label{DA}\eeq
Performing the conformal rescaling as $g_{\mu\nu} = e^{-2D/3}g^c_{\mu\nu}$
for the five-dimensional action (\ref{Ac}) we get
\beq
\frac{L}{16\pi}\int d^5x \sqrt{-g_c}\left( R - (\nabla\phi)^2 -
2(\nabla D)^2 -\frac{e^{-4D/3+2\phi}}{4}H_{+}^2 -
 \frac{e^{-4D/3-2\phi}}{4}H_{-}^2 - \frac{e^{8D/3}}{4}G^2 \right).
\eeq
The five-dimensional black hole with electric charges about both $H_{+}$
and $H_{-}$
\beq
Q_{+} \equiv \frac{1}{8}\int_{S^3} e^{-4D/3+2\phi}*H_{+} ,\;
Q_{-} \equiv \frac{1}{4{\pi}^2}\int_{S^3} e^{-4D/3-2\phi}*H_{-}
\eeq
can also carry the U$(1)$ charge associated with $G$
\beq
 P \equiv \frac{2{\pi}n}{L} = \frac{L}{16\pi}\int_{S^3} e^{8D/3}*G ,
\label{P}\eeq
which is also regarded as the ADM momentum around $S^1$ for the
six-dimensional black string. The parameters $r_{\pm}$ are related to
the charge by $Q^2 \equiv 2Q_{+}Q_{-} = ({\pi}r_{+}r_{-})^2$. This
non-extremal black hole solution is characterized by
\begin{eqnarray}
\phi &=&  \phi_h , \\
e^{-4D/3+2\phi}*H_{+}& =& \frac{4Q_{+}}{\pi^2}\epsilon_3, \;
e^{-4D/3-2\phi}*H_{-} = 2Q_{-}\epsilon_3 , \label{H}\\
ds^2_{5c} &=& - f^{-2/3}\left( 1 - \frac{r_0^2}{\hat{r}^2} \right)dt^2
+ f^{1/3}\left[ \left(1-\frac{r_0^2}{\hat{r}^2}\right)^{-1}d\hat{r}^2
+ \hat{r}^2d\Omega_{3}^2 \right],
\label{dsc}\end{eqnarray}
where $r_0^2 = r_{+}^2 - r_{-}^2$ and $\epsilon_3$ is the volume form on
the unit three-sphere $S^3$. Using the radial coordinate $\hat{r}^2 =
 r^2 -r_{-}^2$, we express $e^{2D}$ as
\beq
e^{2D} = \frac{ 1 + r_0^2\sinh^2\alpha/\hat{r}^2}{ 1 +
r_{-}^2/\hat{r}^2}
\label{D}\eeq
and derive the five-dimensional metric (\ref{dsc}) with $f =
(1 + r_{-}^2/\hat{r}^2)^2(1 + r_0^2\sinh^2\alpha/\hat{r}^2)$
from the six-dimensional metric by the dimensional reduction along
$x_5$ direction through (\ref{ds}), (\ref{dsa}) and $ds_5^2 = e^{-2D/3}
ds_{5c}^2$. The equation of motion for $\phi$
\beq
\nabla^2\phi - \frac{1}{4}e^{-4D/3 +2\phi}H_{+}^2
+ \frac{1}{4}e^{-4D/3 -2\phi}H_{-}^2 = 0
\eeq
is satisfied by a special asymptotic value $\phi_h$ specified by
\beq
e^{2\phi_h} = \frac{2Q_{+}}{\pi^2Q_{-}} ,
\label{Q}\eeq
since $H_{+}, H_{-}$ are simultaneously obtained from (\ref{H}) as
\beq
H^{+}_{t\hat{r}} = \frac{2Q_{-}}{\hat{r}^3( 1 + r_{-}^2/\hat{r}^2)^2},
\; H^{-}_{t\hat{r}} = \frac{4Q_{+}/\pi^2}{\hat{r}^3( 1 +
r_{-}^2/\hat{r}^2)^2} .
\label{HH}\eeq
From the expression of $A_t$ in (\ref{DA}) the U$(1)$ field strength is
determined in a similar form as
\beq
G_{t\hat{r}} =
\frac{r_0^2\sinh2\alpha}{\hat{r}^3( 1 + r_0^2\sinh^2\alpha/\hat{r}^2)^2} ,
\label{G}\eeq
which combines with (\ref{P}) to yield
\beq
P = \frac{\pi L}{8}( r_{+}^2 - r_{-}^2 )\sinh2\alpha.
\eeq
The area $A$ of the event horizon which is located at $\hat{r} = r_0$
corresponding to $r = r_{+}$ can be calculated from the metric (\ref{dsc})
and the entropy of the five-dimensional black hole is given by
\beq
S = \frac{A}{4G_5} = \frac{L}{2}\pi^2r_{+}^2\sqrt{r_{+}^2 - r_{-}^2}
\cosh\alpha .
\label{S}\eeq
This agrees with the entropy of the six-dimensional black string computed
from the metric (\ref{dsa}).

Now in order to take the semiclassical approximation of the path integral
in the Euclid spacetime the above five-dimensional black hole solution
must be analytically continued to Euclidean time. Since the
Einstein-Hilbert action includes a boundary term which is formally infinite,
we must perform an appropriate subtraction \cite{GH,HH}. Recently the
entropy of the Schwarzschild black hole in the presence of a string
instanton has been calculated \cite{EHW}. Following the procedure in
Ref.\cite{EHW}, we will make the subtraction in the computation of the
thermal on-shell action. The subtracted total Euclidean action is given by
\begin{eqnarray}
I & = & I_0 + I_b , \\
I_0 & = & - \frac{L}{16\pi}\int d^5x \sqrt{g_c} ( R - (\nabla\phi)^2
- 2(\nabla D)^2 - \frac{e^{-4D/3 +2\phi}}{4}H_{+}^2 \nonumber \\  & &
- \frac{e^{-4D/3 -2\phi}}{4}H_{-}^2 - \frac{e^{8D/3}}{4}G^2 ) ,\label{IO}
 \\ I_b &=& - \frac{L}{8\pi} \int_{\partial M} \sqrt{h}K + \frac{L}{8\pi}
\int_{(\partial M)_{\infty}} \sqrt{h_0}K_0 ,
\end{eqnarray}
where $K$ is the trace of the extrinsic curvature on the boundary
$\partial M$ of the five-dimensional manifold
$M$, $h$ the determinant of the induced
metric and $K_0, h_0$ the corresponding ones at infinity in flat spacetime.
The  conical singularity at $\hat{r} = r_0$ in (\ref{dsc}) is removed by
requiring the Euclidean time $\tau$ to be identified with period $\beta$
and Bekenstein-Hawking temperature is obtained by
\beq
T \equiv \frac{1}{\beta} =
\frac{\sqrt{r_{+}^2 - r_{-}^2}}{2\pi r_{+}^2\cosh \alpha} ,
\label{T}\eeq
which is compared with that of the six-dimensional black string
$T_6 =\sqrt{r_{+}^2-r_{-}^2}/(2\pi r_{+}^2)$ that is derived from
(\ref{dsa}). Contracting the Einstein equation with the metric tensor
we have
\beq
R = (\nabla \phi)^2 + 2(\nabla D)^2 + \frac{1}{12}( e^{-4D/3 +2\phi}H_{+}^2
 + e^{-4D/3 -2\phi}H_{-}^2 + e^{8D/3}G^2 ) .
\eeq
Substitution of it into (\ref{IO}) yields
\beq
I_0 = \frac{L}{16\pi} \int d^5x \sqrt{g_c} \frac{1}{6}\left(
e^{-4D/3 +2\phi}H_{+}^2 + e^{-4D/3 -2\phi}H_{-}^2 + e^{8D/3}G^2 \right) ,
\label{Ia}\eeq
where the dilaton terms such as $(\nabla \phi)^2, (\nabla D)^2$ have
disappeared. In the Euclid spacetime the U$(1)$ field strengths (\ref{HH}),
(\ref{G}) become pure imaginaries. Therefore putting them into (\ref{Ia})
and combining with (\ref{D}), (\ref{Q}) we find the volume integration
to be simplified into the symmetric two terms as
\beq
I_0 = - \frac{L}{16\pi} \frac{\beta \omega_3}{6}\int_{r_0}^{\infty}
\frac{d\hat{r}}{\hat{r}^3} \left( \frac{32Q_{+}Q_{-}}
{\pi^2(1+ r_{-}^2/\hat{r}^2)^2} + \frac{2r_0^4\sinh^{2}2\alpha}
{(1+ r_0^2\sinh^2\alpha/\hat{r}^2)^2} \right),
\eeq
where the two contributions from $H_{+}^2$ and $H_{-}^2$ have been equal
and $\omega_3$ is the three-dimensional unit-sphere volume. This integration
can be carried out to be
\beq
I_0 = - \frac{L\beta\omega_3}{24\pi}( 2r_{-}^2 + r_0^2\sinh^2\alpha ).
\label{Ib}\eeq

Here to evaluate the boundary term $I_b$ for the non-extremal Euclidean
black hole solution whose boundary is at the infinity, we look into
the asymptotic behavior of the metric for large $\hat{r}$
\beq
ds^2 \sim \left(1 - \frac{2R_0^2 +r_0^2}{\hat{r}^2} \right)d\tau^2 +
\left(1 + \frac{R_0^2 +r_0^2}{\hat{r}^2} \right)d\hat{r}^2
+ ( \hat{r}^2 + R_0^2 )d\Omega_3^2 ,
\eeq
where $R_0^2 = ( 2r_{-}^2 + r_0^2\sinh^2\alpha )/3$. Further the metric
shifted by $\hat{r}^2 + R_0^2 = R^2$ is written as
\beq
ds^2 \sim \left( 1 - \frac{2R_0^2 +r_0^2}{R^2} \right)d\tau^2 +
\left( 1 + \frac{2R_0^2 +r_0^2}{R^2} \right)dR^2 + R^2d\Omega_3^2 .
\eeq
The ADM mass of this solution reads
\beq
M = \frac{3\omega_3}{16\pi G_5}( 2R_0^2 + r_0^2 ) ,
\eeq
which is expressed as
\beq
M = \frac{L\pi}{8}[ 2(r_{+}^2 + r_{-}^2) + \cosh2\alpha
(r_{+}^2 - r_{-}^2) ] .
\eeq
From a spacelike unit vector $n_{\mu} = (0, N, 0, 0, 0)$ normal to the
boundary surface $R = R_{\infty}$, where $N = \sqrt{g_{RR}} =
( 1 + (2R_0^2 + r_0^2)/R^2 )^{1/2}$ is the lapse function, we calculate
the trace of the extrinsic curvature on the boundary $K =
h^{\mu\nu}\nabla_{\mu}n_{\nu} = ( g^{RR}\partial_Rg_{RR} +
\sum_i g^{\theta_i\theta_i}\partial_Rg_{\theta_i\theta_i} )/(2N)$
with the induced metric on the boundary $h_{\mu\nu} = g_{\mu\nu}
- n_{\mu}n_{\nu}$ and obtain
\beq
\frac{L}{8\pi}\int_{\partial M} \sqrt{h}K = \frac{L\beta \omega_3}
{16\pi} \left[ 6R_{\infty}^2\left( 1 - \frac{2R_0^2 + r_0^2}{R_{\infty}^2}
\right) + 2(2R_0^2 + r_0^2) \right] .
\eeq
The subtracted term is derived for the flat metric
\beq
ds^2 \sim \left( 1 - \frac{2R_0^2 + r_0^2 }{R^2} \right)d\tau^2 + dR^2
+ R^2d\Omega_3^2
\eeq
in a similar fashion to be
\beq
\frac{L}{8\pi}\int_{(\partial M)_{\infty}} \sqrt{h_0}K_0 =
\frac{L\beta \omega_3}{16\pi} 6R_{\infty}^2
\left( 1 - \frac{ 2R_0^2 + r_0^2 }{2R_{\infty}^2} \right) .
\eeq
Gathering together we have $I_b = \beta M/3$. Since an interesting relation
$ST = L\pi r_0^2/4$ makes $I_0$ (\ref{Ib}) expressed as
\beq
I_0 = \beta\left( - \frac{M}{3} + \frac{1}{2}ST \right) ,
\label{Ic}\eeq
we arrive at the Euclidean on-shell action
\beq
I = \frac{1}{2}S .
\eeq
The expression $I_0$ (\ref{Ib}) is alternatively described by
\beq
I_0 = - \frac{\beta}{3} ( Q\Phi_Q + P\Phi_P ) ,
\label{Id}\eeq
where $\Phi_Q = LQ/2r_{+}^2$ is the electric potential on the horizon
in five dimensions \cite{GR,GM} and $\Phi_P = 4P/L(r_0\cosh\alpha)^2$
may be regarded as the electric potential associated with $G$.
To consider the thermodynamics of the five-dimensional
black hole we define the Gibbs free energy
\beq
W = M - TS - Q\Phi_Q - P\Phi_P = - T\log Z .
\eeq
At finite temperature the classical Euclidean action $I$ can be related
approximately to a thermal partition function $Z$ as $Z \sim e^{-I}$,
which yields
\beq
W = IT = \frac{1}{2}ST .
\eeq
Combining them we obtain
\beq
M = \frac{3}{2}ST + Q\Phi_Q + P\Phi_P ,
\label{M}\eeq
which is also seen from (\ref{Ic}) and (\ref{Id}). In Ref.\cite{GM} the
electrically singly-charged $d$-dimensional black hole solution in the
Maxwell dilaton Einstein theory was constructed and the generalized Smarr
formula $M = ((d-2)/(d-3))ST + Q\Phi_Q$ was presented. Our one result
(\ref{M}) is considered to exhibit the generalized Smarr formula in five
dimensions, which corresponds to the Smarr formula $M = 2ST + Q\Phi_Q$
for the four-dimensional electrically singly-charged dilatonic black
hole \cite{GR}. The other result that the Euclidean on-shell action is
one half of the entropy in five dimensions is compared with the
four-dimensional case where it equals to the entropy itself
\cite{KOP,GR}. As seen above the two results are consistently related
to each other.

Now let us consider first the near-extremal black hole by parametrizing
the left-moving and right-moving oscillations of a fundamental string
around $S^1$ as
\beq
n_L \equiv \frac{e^{-2\alpha}}{2\sinh 2\alpha}n \sim n', \;
n_R \equiv \frac{e^{2\alpha}}{2\sinh 2\alpha}n \sim n + n'
\label{n}\eeq
to keep the total charge $n = n_R - n_L$ fixed in the large $\alpha$
region with small $r_0$. The mass and entropy of it are expressed as
$ M = LQ/2 + (n + 2n')/R, S = \sqrt{2}\pi Q (\sqrt{n + n'} + \sqrt{n'})$
with $L =2\pi R$. This near-extremal entropy has been understood from
a counting of the underlying microscopic degree of freedom by using
the weak-coupling D-brane description \cite{HS,CM,HMS}. Using these
expressions we can demonstrate a thermal relation $dM = TdS$
with $T = 2\sqrt{2n'(n +n')}/(\pi QR(\sqrt{n +n'} + \sqrt{n'}))$.
This Bekenstein-Hawking temperature is further described in terms of
the right-moving temperature $T_R = \sqrt{2(n +n')}/\pi QR$ and
the left-moving one $T_L = \sqrt{2n'}/\pi QR$ as a harmonic mean
$2/T = 1/T_R + 1/T_L$ \cite{MS2}. The specific heat at constnat charges
$Q, P$ is evaluated as
\beq
C = \left( \frac{\partial M}{\partial T} \right)_{Q,P} =
\frac{S}{(n + 2n')/\sqrt{n'(n +n')} - 1} .
\label{C}\eeq
This tells us that near extremality the specific heat is positive and
when we approach the extremality it goes to zero.

We will look more closely at the specific heat far off extremality.
In doing so, we work on the non-extremal black hole solution parametrized
by $r_+, r_-$ and $\alpha$. Instead choosing $Q, P$ and $\alpha$ as
the independent parameters we can express $r_\pm, M$ as
\beq
r_{\pm}^2 = \pm \frac{4P}{\pi L\sinh 2\alpha} + \sqrt{D}
\eeq
with $D = (Q/\pi)^2 + (4P/\:\pi L\sinh 2\alpha)^2$,
\beq
M = \frac{L\pi}{2}\left( \sqrt{D} + \frac{2P}{\pi L\tanh 2\alpha} \right).
\label{Ma}\eeq
Taking the derivative of (\ref{S}) expressed in terms of $Q, P, \alpha$ as
\beq
S = \frac{L{\pi}^2}{2}\sqrt{\frac{4P}{\pi L\tanh \alpha}}
\left(\sqrt{D} + \frac{4P}{\pi L\sinh 2\alpha}\right)
\eeq
with respect to $\alpha$ at constant $Q, P$ we can obtain a factorized
form
\beq
\frac{\partial S}{\partial \alpha} = - \frac{L{\pi}^2}{2}
\sqrt{\frac{4P}{\pi L\tanh \alpha}}\frac{(\sqrt{D}\sinh 2\alpha +
\frac{4P}{\pi L})(\sqrt{D}\sinh 2\alpha + \frac{8P}{\pi L}
\cosh 2\alpha)}{\sqrt{D}\sinh^3 2\alpha} ,
\eeq
which produces $dM =TdS$ again with
\beq
T = \frac{\sqrt{4P\tanh \alpha/\, \pi L}}{\pi(4P/\, \pi L +
\sqrt{D}\sinh 2\alpha)},
\label{Ta}\eeq
which is given from (\ref{T}). The specific heat
is also computed from (\ref{Ma}), (\ref{Ta}) to be
\beq
C^{-1} = \frac{1}{S}\left( \frac{2\sqrt{D}\cosh 2\alpha }{\sqrt{D} +
8P/(\pi L\tanh 2\alpha)} - 1 \right) ,
\eeq
which is in a similar form to (\ref{C}).
In the large $\alpha$ region the specific heat reduces to the
near-extremal positive one (\ref{C}) through $e^{2\alpha} \sim
\sqrt{(n +n')/n'}$ which is given from (\ref{n}), while in the
small $\alpha$ region it turns out to be negative. Therefore the specific
heat blows up at the point $\alpha = \alpha_0$ provided by the solution of
\beq
f(x) \equiv Q^2(x^2 - 1)(2x - 1)^2 - \left(\frac{4P}{L}\right)^2
(4x -1) = 0
\eeq
with $f(1)< 0$ for $x = \cosh 2\alpha \geq 1$. At the particular black hole
mass specified by $\alpha_0$ there happens
a phase transition. For the zero charge $Q = 0$
with $P \neq 0$ we see that the specific heat becomes always negative
\beq
C = - L\pi^2 \left(\frac{4P}{\pi L}\right)^{3/2} \frac{ 1 +
2\cosh 2\alpha}{\sinh 2\alpha \sqrt{\tanh \alpha}} ,
\label{Ca}\eeq
whose expression has already appeared in the above small $\alpha$ region
with nonzero $Q, P$. This sign is characteristic of the Kaluza-Klein
black holes \cite{AG}. On the other hand in the small $P$ region with
nonzero $Q$ and finite $\alpha$ the specific heat takes small positive
value. Therefore we can say that the existence of both $Q$ and $P$ charges
is responsible for the appearance of the singularity. This
singularity is compared with that in the four-dimensional U$(1)^2$
dilaton black hole, where the singularity appears only when it has
both electric and magnetic charges \cite{KLOP}. Similarly in the
four-dimensional electrically singly-charged black hole paramertrized
by $a$, which characterizes the coupling of the dilaton to the gauge
field, the phase transition was found only for $a^2 < 1$
\cite{GR,GM}. The specific heats for the $a = 1$ stringy black hole and
the $a = \sqrt{3}$ Kaluza-Klein black hole are always negative, which
are considered to correspond to the disappearances of singularities for
the U$(1)^2$ dilaton black hole with only electric charge \cite{KLOP}
and our $Q = 0, P \neq 0$ case in (\ref{Ca}) respectively.

In conclusion we have made the Euclidean semi-classical path-integral
analysis of the thermodynamics of the five-dimensional non-extremal black
hole with multiple electric charges. We have performed the explicit
calculation of the on-shell action with the suitably subtracted boundary
term and found that it is one half of the black hole entropy. This
semi-classical result has been shown to be consistently connected with the
generalized Smarr formula for the five-dimensional black hole mass.
We expect such connections to appear in the other dimensions.
We have carried out the computation of the specific heat of the
non-extremal black hole, which was guided by that of the near-extremal
one, and observed that there exists a phase transition. It would be
interesting to understand this phase transition presented by the
strong-coupling black hole description from the alternative
weak-coupling D-brane description.

\end{document}